\documentclass[aps,prl,twocolumn,preprintnumbers,superscriptaddress,nofootinbib]{revtex4-1}
\usepackage[utf8]{inputenc}
\usepackage{psfrag}
\usepackage{amssymb}
\usepackage{amsmath}
\usepackage{epstopdf}
\usepackage{color}
\usepackage{graphicx}
\usepackage{subcaption}
\usepackage{booktabs}
\usepackage{soul}
\usepackage{ragged2e}

\usepackage{lmodern}
\usepackage{hyperref}
\hypersetup{colorlinks=true,linkcolor=red,anchorcolor=blue,citecolor=blue, filecolor=blue,urlcolor=red,bookmarksnumbered=true,
pdfview=FitB
}

\begin{document}
\title{Origin of Tensor Polarized Structure of the Deuteron}
\author{Satvir~Kaur}
\email{satvir@impcas.ac.cn}
\affiliation{Institute of Modern Physics, Chinese Academy of Sciences, Lanzhou 730000, China}
\affiliation{University of Chinese Academy of Sciences, Beijing 100049, China}
\author{Chandan~Mondal}
\email{mondal@impcas.ac.cn} 
\affiliation{Institute of Modern Physics, Chinese Academy of Sciences, Lanzhou 730000, China}
\affiliation{University of Chinese Academy of Sciences, Beijing 100049, China}
\author{Xingbo~Zhao}
\email{xbzhao@impcas.ac.cn} 
\affiliation{Institute of Modern Physics, Chinese Academy of Sciences, Lanzhou 730000, China}
\affiliation{University of Chinese Academy of Sciences, Beijing 100049, China}
\affiliation{Advanced Energy Science and Technology Guangdong Laboratory, Huizhou, Guangdong 516000, China}
\author{Chueng-Ryong~Ji}
\email{crji@ncsu.edu} 
\affiliation{Department of Physics and Astronomy, North Carolina State University,
Raleigh, North Carolina 27695-8202, USA}

\begin{abstract}
The microscopic origin of the deuteron tensor polarized structure function $b_1$ remains unresolved. Conventional nuclear models attribute a nonzero $b_1$ mainly to the $D$-wave component of the deuteron wave function, but generally fail to reproduce the magnitude and momentum-dependence observed by the HERMES experiment. Using a QCD-based light-front framework with degrees of freedom beyond the conventional nucleonic picture, in particular including hidden-color components and relativistic spin--momentum correlations, we find that a small but nonzero estimated hidden-color admixture can be sufficient to generate a substantial tensor response ($b_1(x)$),
while it is essentially insensitive to the $D$-wave component. The physical $D$-wave probability of $0.85\%$, fixed by the empirical quadrupole moment, changes $b_1$ by only $1.7\%$, with the dominant contribution arising from the colored-cluster $S$-wave sector. These results demonstrate that the tensor-polarized structure of the deuteron is driven primarily by hidden-color dynamics rather than by the $D$-wave component.
\end{abstract}

\maketitle

{\it Introduction}.---The first measurement of the deuteron tensor polarized
structure function $b_1$ by the HERMES Collaboration revealed a nonzero
signal~\cite{HERMES:2005pon}. Its microscopic origin remains a
longstanding puzzle within the conventional description of the deuteron
as a weakly bound proton--neutron system. As a leading-twist observable
of a spin-$1$ target, $b_1$ measures the difference between the
quark distributions in the helicity-zero deuteron state and their
average in the helicity-$\pm1$ states, thereby providing direct access
to the tensor partonic structure of the deuteron~\cite{Hoodbhoy:1988am}.

In the conventional nuclear description, the deuteron wave function is
dominated by the ${}^{3}S_1$ state, with a small ${}^{3}D_1$ component
generated by the nucleon--nucleon tensor force. Within nucleonic
convolution models, $b_1$ vanishes for a pure $S$-wave deuteron, while
the $D$ wave generates a nonzero contribution through $S$--$D$
interference and the pure $D$-wave term~\cite{Hoodbhoy:1988am,Khan:1991qk,Cosyn:2017fbo,Cao:2026wxo}. This has led to the
common association of a nonzero $b_1$ with the $D$-wave component.
However, conventional calculations generally predict a smaller
magnitude and fail to reproduce the trend indicated by the HERMES data~\cite{Khan:1991qk,Cosyn:2017fbo}. Likewise, a recent convolution analysis of
tensor-polarized parton distributions finds substantial
differences from those inferred from the HERMES measurement~\cite{Kumano:2026ipc}.

Several mechanisms beyond the standard nucleonic convolution picture
have been considered. Coherent nuclear shadowing can contribute
significantly at small Bjorken $x$~\cite{Edelmann:1997ik}, while pionic
degrees of freedom remain small at moderate and large $x$~\cite{Miller:2013hla}. Phenomenological analyses have also indicated
the possible presence of tensor-polarized sea-quark distributions~\cite{Kumano:2010vz}. Non-nucleonic six-quark configurations,
particularly hidden-color components, provide another possible source
of the observed nonzero $b_1$~\cite{Miller:2013hla,Ji:2014cda,Bakker:2014cua}.


Clarifying these mechanisms is particularly timely in view of planned and proposed measurements of $b_1$ at Jefferson Lab~\cite{ProposalJLab2023}, tensor-polarized deuteron observables at future Electron--Ion Colliders~\cite{Kalantarians:2014eda,Anderle:2021wcy}, and Drell--Yan experiments at Fermilab~\cite{Kumano:2016ude} and NICA~\cite{Savin:2016arw}. Recent studies of tensor-polarized deuteron transverse-momentum-dependent parton distributions (TMDs) in semi-inclusive deep-inelastic scattering (SIDIS) further motivate investigations of related structure functions, as both probe the same underlying quark correlations from complementary perspectives~\cite{Kumano:2024fpr,Zhao:2025vol,Poudel:2025nof,Poudel:2025tac}. In this context, our study of $b_1$ provides a complementary probe of these correlations, with particular emphasis on hidden-color dynamics.

In this Letter, we elucidate the microscopic origin of $b_1$ within a QCD-based light-front (LF) framework\footnote{The LF variables are defined as
$a^\pm = a^0 \pm a^3$ and ${
\vec{a}}_\perp = (a^1, a^2)$.} that incorporates hidden-color, relativistic spin, and orbital dynamics. We include both ${}^{3}S_1$ and ${}^{3}D_1$ components, with the $D$-wave probability $P_D=0.85\%$ fixed independently from the empirical quadrupole moment. The maximum absolute difference between the $S$- and $S+D$-wave results
for $b_1(x)$, normalized to the maximum magnitude of the $S$-wave
result, is only $1.7\%$. This small difference indicates that hidden-color dynamics, rather than the conventional $D$-wave mechanism, dominates the tensor polarized structure function of the deuteron.

{\it Hidden-color dynamics}.---At hadronic resolution, the deuteron is
conventionally described as a weakly bound system of two color-singlet
nucleons. At the quark level, however, its six quarks can couple to an
overall color singlet through both singlet--singlet and hidden-color
channels. We write this structure schematically as
\begin{equation}
|\mathcal{D}\rangle
=
c_{11}\,|\psi_{1\otimes1}\rangle
+
c_{88}\,|\psi_{8\otimes8}\rangle;
\qquad
|c_{11}|^2+|c_{88}|^2=1,
\label{eq:color_decomposition}
\end{equation}
where $|\psi_{1\otimes1}\rangle$ represents the conventional
color-singlet nucleon--nucleon channel. The state
$|\psi_{8\otimes8}\rangle$ collectively represents configurations in
which two colored subclusters are coupled to an overall color singlet.
These configurations are referred to as ``hidden color" because the
individual colored clusters cannot appear as asymptotic physical
states but entangled in their color degrees of freedom with the deuteron. They are expected to become increasingly relevant when the
deuteron is resolved at short distances or high resolutions
~\cite{Ji:2014cda,Bakker:2014cua}.

Orbital and color dynamics represent distinct aspects of the deuteron
wave function. The $D$-wave describes its nonspherical orbital
structure and is essential for the quadrupole moment, whereas hidden
color describes non-nucleonic couplings among its six quarks. Including
both effects allows us to test whether $b_1$ is controlled by orbital
deformation or by the internal color dynamics.

The interplay between hidden-color and
orbital dynamics can be studied using different QCD-motivated
descriptions. Here, we employ a LF realization in which
the hidden-color sector is represented by two effective colored
clusters coupled to an overall color-singlet state
~\cite{Kaur:2025css}. This construction provides one concrete way to
incorporate non-nucleonic color degrees of freedom while retaining a
relativistic description of the internal spin and orbital dynamics.

The corresponding light-front wave function (LFWF) depends on the longitudinal momentum fraction $z$, relative transverse momentum $\boldsymbol{k}_{\perp}$, deuteron LF spin projection $\Lambda$, and cluster helicities $\lambda_1$ and $\lambda_2$, and can be decomposed as
\begin{equation}
\Psi^{\Lambda}_{\lbrace\lambda_1\lambda_2\rbrace}
(z,\boldsymbol{k}_{\perp})
=
\sum_{\substack{L,L_z\\S,S_z}}
C^{\,1\Lambda}_{L L_z,S S_z}\,
\widetilde{\psi}_{L L_z}
(z,\boldsymbol{k}_{\perp})\,
\mathcal{S}^{S S_z}_{\lambda_1\lambda_2}
(z,\boldsymbol{k}_{\perp}),
\label{eq:deuteron_LFWF}
\end{equation}
where
$C^{\,1\Lambda}_{L L_z,S S_z}
=
\langle L L_z,S S_z\vert1\Lambda\rangle
$ is the Clebsch–Gordan which
couples the orbital angular momentum $(L,L_z)$ and total cluster spin
$(S,S_z)$ to the deuteron spin $J=1$. We retain the
${}^{3}S_1$ and ${}^{3}D_1$ channels, corresponding to
$(L,S)=(0,1)$ and $(2,1)$, respectively. The function
$\widetilde{\psi}_{L L_z}$ describes the relative cluster motion, while
$\mathcal{S}^{S S_z}_{\lambda_1\lambda_2}$ is obtained by applying the
Melosh rotations to the canonical cluster-spin state~\cite{Melosh:1974cu,Ji:1992yf}.

The complete deuteron LFWF is expressed as
\begin{equation}
\Psi^\Lambda
=
a_S\Psi_S^\Lambda+a_D\Psi_D^\Lambda;
\qquad
|a_S|^2+|a_D|^2=1,
\label{eq:SD_decomposition}
\end{equation}
where $\Psi_S^\Lambda$ and $\Psi_D^\Lambda$ denote the
${}^{3}S_1$ and ${}^{3}D_1$ components with the probabilities $|a_S|^2$ and  $|a_D|^2$, respectively.

In the mixed longitudinal-momentum and transverse-coordinate
representation, the wave function factorizes as~\cite{Brodsky:2014yha}
\begin{equation}
\psi_{L L_z}(z,\boldsymbol{b}_{\perp})
=
\frac{\phi_L(\zeta)}{\sqrt{2\pi\zeta}}\,
e^{iL_z\theta}\,
X(z),
\qquad
\zeta=\sqrt{z(1-z)}\,b_\perp,
\label{eq:coordinate_LFWF}
\end{equation}
where $L=|L_z|$, $\boldsymbol{b}_{\perp}$ is the transverse separation
between the clusters, $b_\perp=|\boldsymbol{b}_{\perp}|$, and $\theta$
is the azimuthal angle of $\boldsymbol{b}_{\perp}$. The transverse radial
mode $\phi_L(\zeta)$ is obtained by solving the LF holographic
Schr\"odinger equation~\cite{Brodsky:2014yha}, while the longitudinal
mode $X(z)$ is determined from the 't~Hooft equation
~\cite{tHooft:1974pnl}. The factor $e^{iL_z\theta}$ carries the
azimuthal dependence associated with the orbital-angular-momentum
projection $L_z$.

The momentum-space wave function follows from the transverse Fourier
transform
\begin{equation}
\widetilde{\psi}_{L L_z}
(z,\boldsymbol{k}_{\perp})
=
\int\frac{d^2\boldsymbol{b}_{\perp}}{(2\pi)^2}\,
e^{-i\boldsymbol{k}_{\perp}\cdot\boldsymbol{b}_{\perp}}\,
\psi_{L L_z}(z,\boldsymbol{b}_{\perp}),
\label{eq:momentum_LFWF}
\end{equation}
where $\boldsymbol{k}_{\perp}$ is conjugate to
$\boldsymbol{b}_{\perp}$. 
The LFWF is obtained by solving the  LF
Schr\"odinger-like equations for the transverse and longitudinal
dynamics, as detailed in Ref.~\cite{Kaur:2025css}. The deuteron mass
eigenvalue and the corresponding LFWF are determined by three
parameters: the effective cluster mass $m_\mathcal{C}$, the
longitudinal confinement strength $g$, and the transverse confinement
scale $\kappa$. The parameter $g$ enters the 't~Hooft equation for the
longitudinal mode, whereas $\kappa$ determines the LF holographic
potential for the transverse mode. Requiring the physical deuteron mass
correlates these three parameters and fixes the corresponding LFWF.

The resolution of the effective two-cluster configuration is
characterized by the probing energy\footnote{In Ref.~\cite{Kaur:2025css}, the quantity
$E_{\rm p}=M_D-2m_\mathcal{C}$ was referred to as the binding energy.
We instead call it the probing energy to distinguish it from the
physical deuteron binding energy $\simeq2.2~\mathrm{MeV}$. The definition and numerical
quantity are unchanged; $E_{\rm p}$ characterizes the resolution of
the effective colored-cluster description.}
$
E_{\rm p}=M_D-2m_{\mathcal{C}}$,
where $M_D$ is the deuteron mass. Small $E_{\rm p}$ corresponds to the weakly bound
nucleonic regime, whereas increasing $E_{\rm p}$ probes
shorter-distance dynamics for which colored-cluster configurations can
contribute. Although $E_{\rm p}$ does not directly quantify the
hidden-color content of the deuteron wave function, it serves as an
effective probe of these short-distance color dynamics.

\begin{figure}[t]
\centering
\includegraphics[width=\columnwidth]{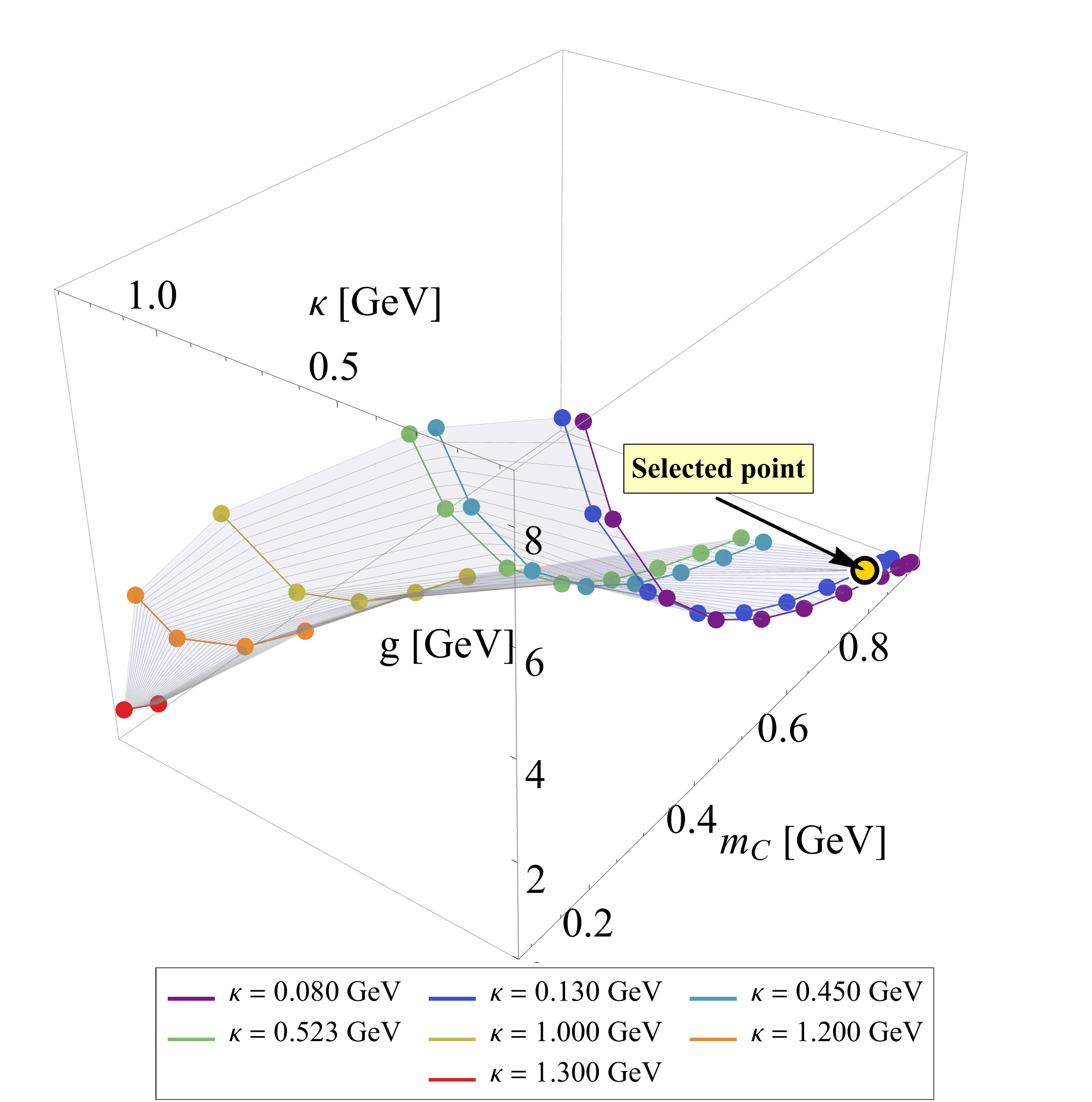}
\caption{\justifying Combinations of the effective cluster mass $m_\mathcal{C}$ and the
longitudinal and transverse confinement strengths $g$ and $\kappa$, respectively that
reproduce the deuteron mass. The highlighted point denotes the parameter
set used in this work, $\{m_\mathcal{C},g,\kappa\}
=\{0.838,0.50,0.13\}~\mathrm{GeV}$, corresponding to
$E_{\rm p}\simeq200~\mathrm{MeV}$.}
\label{fig:3D-mC-g-kappa}
\end{figure}

Figure~\ref{fig:3D-mC-g-kappa} displays the correlated combinations of
$m_\mathcal{C}$, $g$, and $\kappa$ that satisfy the deuteron mass
eigenvalue condition. Along the solutions shown, increasing
$m_\mathcal{C}$ reduces the confinement strengths required to reproduce
$M_D$, whereas decreasing $m_\mathcal{C}$ requires larger values of
both the effective confinement scales. This trend is consistent with the expected color structure of the
deuteron at different resolutions. In the formal chiral limit
$m_\mathcal{C}\rightarrow0$, the system becomes more compact and
strongly interacting; color recoupling then implies an octet--octet
probability of about $80\%$
~\cite{Ji:2014cda,Bakker:2014cua}. Conversely, as
$m_\mathcal{C}\rightarrow m_N$, the required interactions weaken and
the system approaches the conventional weakly bound singlet--singlet
nucleonic configuration, for which $P_{88}=|c_{88}|^2\rightarrow0$.

For the consistency of our analysis approximating the cluster quark PDFs as the unpolarized nucleon PDFs from the NNPDF fits, we use
$m_\mathcal{C}=0.838~\mathrm{GeV}$,
$g=0.50~\mathrm{GeV}$, and $\kappa=0.13~\mathrm{GeV}$. This set
corresponds to the highlighted point and to
$E_{\rm p}\simeq200~\mathrm{MeV}$~\cite{Kaur:2025css}. The selected point, corresponding to
$E_{\rm p}\simeq200~\mathrm{MeV}$, stays close to the nucleonic limit
and is well separated from the formal chiral limit (as shown in Fig.~\ref{fig:3D-mC-g-kappa}). Although the mass
eigenvalue condition does not directly determine the color amplitudes
$c_{11}$ and $c_{88}$, the hidden-color content may be estimated
indirectly by interpolating between the two limiting regimes. Assuming
a linear interpolation in $m_\mathcal{C}$ between the compact six-quark limit,
$P_{88}=80\%$ at $m_\mathcal{C}=0$, and the nucleonic limit,
$P_{88}=0$ at $m_\mathcal{C}=m_N$, we obtain
$
P_{88}^{\rm est}
=
0.8\left(1-\frac{m_\mathcal{C}}{m_N}\right)
\simeq8.5\%.$
This provides a qualitative estimate of the hidden-color admixture
rather than a probability obtained through an explicit color
projection. Within this qualitative interpolation, we estimate that the deuteron
contains a hidden-color admixture of less than $10\%$.

To fix the orbital admixture, we require the full $S+D$ wave function to reproduce the empirical deuteron quadrupole moment, obtaining $P_D=|a_D|^2=0.85\%$. This relatively small $D$-wave probability should be contrasted with the $(\sim4-6\%)$ $D$-wave admixture~\cite{Machleidt:2000ge,Wiringa:1994wb,Stoks:1994wp} commonly found in conventional descriptions of the deuteron as a weakly bound nucleon--nucleon ($NN$) system. In the present approach, however, the deuteron is described in terms of strongly interacting color clusters rather than as a shallow $NN$ bound state. In particular, the inclusion of hidden-color components, even though their overall probability is less than $10\%$, modifies the internal dynamics and leads to a substantially smaller $D$-wave admixture. Thus, the small value of $P_D$ is a consequence of the underlying cluster structure of the deuteron with the entangled color degrees of freedom rather than an additional phenomenological assumption. 

Despite its small probability, the $D$-wave remains important for observables that are particularly sensitive to the orbital structure. In particular, the quadrupole moment receives a strong contribution from the interference between the $S$- and $D$-wave components. 
Consequently, the quadrupole moment is highly sensitive to the $D$ wave:
the $S$-wave result, $Q_D^{(S)}=0.079~\mathrm{GeV}^{-2}$, is brought to
the empirical value
$Q_D^{\rm exp}=7.34~\mathrm{GeV}^{-2}$
~\cite{Garcon:2001sz} by the $0.85\%$ $D$-wave admixture. We therefore use this independently determined value without further adjustment when calculating $b_1$.

In contrast, the small admixture has only a minor effect on the charge
and magnetic observables. The charge radius changes from
$2.17\pm0.20~\mathrm{fm}$ to $2.21\pm0.21~\mathrm{fm}$, to be compared
with the experimental value
$r_C^{\rm exp}=2.130\pm0.003\pm0.009~\mathrm{fm}$~\cite{Sick:1998cvq}. The quoted theoretical uncertainties are obtained by varying the
parameters $m_\mathcal{C}$, $g$, and $\kappa$ by $10\%$ around their
central values. The magnetic radius remains
$1.98\pm0.19~\mathrm{fm}$, consistent with the extracted value
$r_M^{\rm exp}=1.90\pm0.14~\mathrm{fm}$~\cite{Afanasev:1998}, while the magnetic moment
changes only from $0.84$ to $0.82$, close to the empirical value
$\mu_D^{\rm exp}=0.85$~\cite{Garcon:2001sz}. The unpolarized and
helicity-dependent distributions are likewise essentially unchanged~\cite{Kaur:2025css}.

{\it Tensor polarized structure function}.---Using the well-established convolution framework, implemented for
effective clusters in Ref.~\cite{Kaur:2025css}, we evaluate the quark
contribution to the tensor structure function as
\begin{equation}
b_1(x,Q^2)
=
\frac{1}{2}\sum_{q,N} e_q^2
\int_x^1\frac{dz}{z}\,
f_{1LL}(z)\,
q^N\!\left(\frac{x}{z},Q^2\right),
\label{eq:b1_convolution}
\end{equation}
where $N$ labels the two effective clusters, $z$ is their longitudinal
momentum fraction, and $e_q$ is the quark charge in units of the
positron charge. The tensor-polarized longitudinal momentum
distribution is
$f_{1LL}(z)=2P_\uparrow^0(z)
-[P_\uparrow^{+1}(z)+P_\uparrow^{-1}(z)]$,
where $P_\uparrow^\Lambda(z)$ denotes the density of a cluster with
helicity $+1/2$ in a deuteron with LF helicity $\Lambda$, obtained
from the LFWF by integrating over transverse momentum and summing
over the spectator helicity. The cluster quark PDFs
$q^N$ are approximated by unpolarized nucleon PDFs from
the NNPDF fits at $Q^2\simeq5~\mathrm{GeV}^2$, as in
Ref.~\cite{Kaur:2025css}.

\begin{figure}[thb!]
    \centering
    \includegraphics[width=\columnwidth]{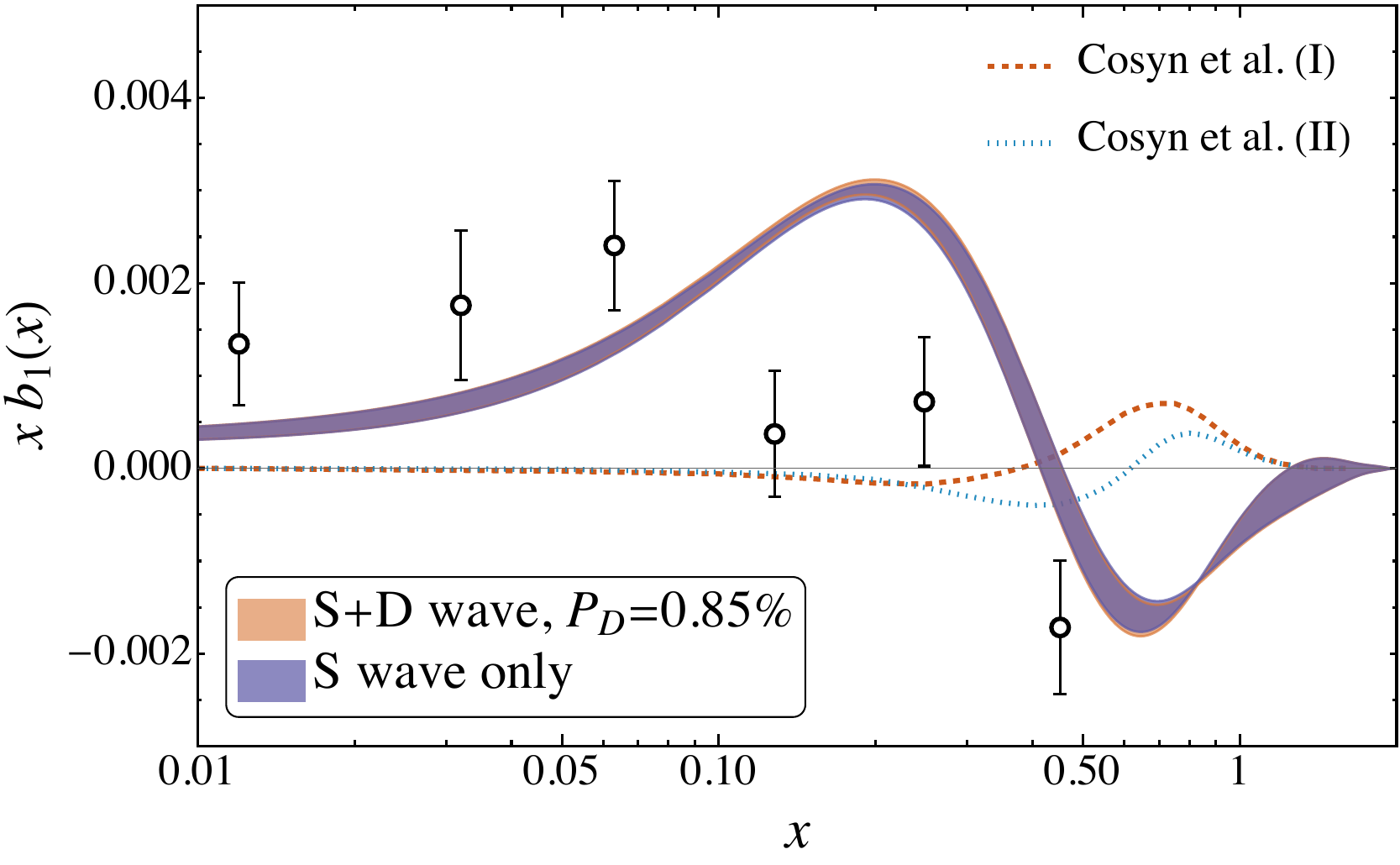}
    \caption{\justifying Deuteron tensor structure function $x b_1(x)$ at
$Q^2=5~\mathrm{GeV}^2$. The orange-shaded band shows the full $S+D$
result with $P_D=0.85\%$, while the purple-shaded band denotes the pure
$S$-wave result. The bands reflect the $10\%$ variation of
$m_\mathcal{C}$, $g$, and $\kappa$ around their central values. The dashed and dotted curves show the
conventional nucleonic calculations of Ref.~\cite{Cosyn:2017fbo}:
basic convolution and the virtual-nucleon approximation,
respectively. Open circles with error bars represent the HERMES
data~\cite{HERMES:2005pon}. 
}
    \label{fig:b1}
\end{figure}



As shown in Fig.~\ref{fig:b1}, the $S$- and $S+D$-wave results remain
close throughout the $x$ range considered. Their maximum absolute
difference, normalized to the maximum magnitude of the $S$-wave result,
is only $1.7\%$. The description of the HERMES data
~\cite{HERMES:2005pon} is therefore essentially unchanged by the
inclusion of the $D$ wave.

Figure~\ref{fig:b1} also compares our result with the two conventional
nucleonic calculations of Ref.~\cite{Cosyn:2017fbo}, obtained using the
basic convolution approach and the virtual-nucleon approximation. In
the ordinary nucleonic picture, a pure $S$-wave deuteron gives
$b_1=0$; a nonzero tensor response arises from $S$--$D$ interference and
the pure $D$-wave contribution. Their predictions nevertheless differ
from the HERMES data in magnitude and $x$ dependence. By contrast, the
colored-cluster $S$-wave sector already produces a nonzero result
consistent with the data. The $D$ wave then provides only a small
correction, despite being required by the quadrupole moment. This
comparison identifies hidden-color dynamics as the dominant source of
the tensor response.

{\it Conclusion and outlook}.---Our results identify a source of the deuteron
tensor structure function beyond the conventional mechanism.
Within the present relativistic light-front description, hidden-color
degrees of freedom produce a nonzero $b_1$ already in the $S$-wave
sector, consistent with the HERMES data. These results indicate that a small but nonzero estimated hidden-color admixture,
below $10\%$, can be sufficient to generate a substantial tensor
response. Adding a $D$-wave component with probability $P_D=0.85\%$,
fixed independently by the empirical quadrupole moment, changes $b_1$
only slightly: the maximum absolute difference is $1.7\%$ of the
maximum magnitude of the $S$-wave result. The same $D$-wave admixture brings the quadrupole moment into agreement
with its empirical value,
while leaving the charge and magnetic observables and the unpolarized
and helicity-dependent distributions essentially unchanged.

The $D$ wave needed to reproduce the quadrupole moment therefore does
not drive the tensor partonic response. Instead, the persistence of
$b_1$ in the colored-cluster $S$-wave sector points to non-nucleonic
color dynamics as an important ingredient in its microscopic origin.
Precision measurements of $b_1$ can test this interpretation and help
distinguish the roles of orbital motion and color dynamics in the
deuteron.

We are aware of ongoing efforts to measure the transverse-momentum
structure of the deuteron through tensor-polarized SIDIS measurements at
CLAS12~\cite{Poudel:2025tac}. The available projections for these
observables remain approximate. In our forthcoming study, we will explore how
the hidden-color dynamics essential for the inclusive tensor response
manifest themselves in the transverse structure of the deuteron. Our preliminary results 
indicate that hidden-color dynamics 
are likewise essential for describing the transverse
tensor-polarized distributions. 



{\it Acknowledgements}.---We thank Siqi Xu,
Gerald A. Miller and James P. Vary for insightful discussions. XZ is supported by National Natural Science Foundation of China, Grant No. 12375143, by Key Research Program of Frontier Sciences, Chinese Academy of
Sciences, Grant No. ZDBS-LY-7020, by international partnership program of the Chinese Academy of Sciences, Grant
No. 016GJHZ2022103FN, by the Strategic Priority Research Program of the Chinese Academy of Sciences, Grant No.
XDB34000000. CRJ is
supported in part by the U.S. Department of Energy (Grant
No. DE-FG02-03ER41260) and acknowledges the National
Energy Research Scientific Computing Center (NERSC)
supported by the Office of Science of the U.S. Department of
Energy under Contract No. DE-AC02-05CH11231.

\bibliography{ref.bib}
\end{document}